\documentclass{article}
\usepackage[utf8]{inputenc}
\usepackage[english]{babel}

\usepackage{blindtext}
\usepackage{hyperref}

\title{Meta-research on COVID-19\\
\textit{An overview of the early trends}}
\author{Giovanni Colavizza\footnote{For questions, suggestions and to signal omissions, please write to: \url{g.colavizza@uva.nl}.} \\
\textit{University of Amsterdam}}
\date{}

\begin{document}

\maketitle

\begin{abstract}
    COVID-19 is having a dramatic impact on research and researchers. The pandemic has underlined the severity of known challenges in research and surfaced new ones, but also accelerated the adoption of innovations and manifested new opportunities. This review considers early trends emerging from meta-research on COVID-19. In particular, it focuses on the following topics: i) mapping COVID-19 research; ii) data and machine learning; iii) research practices including open access and open data, reviewing, publishing and funding; iv) communicating research to the public; v) the impact of COVID-19 on researchers, in particular with respect to gender and career trajectories. This overview finds that most early meta-research on COVID-19 has been reactive and focused on short-term questions, while more recently a shift to consider the long-term consequences of COVID-19 is taking place. Based on these findings, the author speculates that some aspects of doing research during COVID-19 are more likely to persist than others. These include: the shift to virtual for academic events such as conferences; the use of openly accessible pre-prints; the `datafication' of scholarly literature and consequent broader adoption of machine learning in science communication; the public visibility of research and researchers on social and online media.
    
    %will start with an overview of the volume and pace of COVID-19-related research contributions which have been made before and during the pandemic, highlighting, in particular, the variety of disciplinary perspectives which came to the fore. Special attention will be given to an assessment of the proliferation of tools to support COVID-19 research, including search engines. The chapter will then discuss the impact of COVID-19 on key research practices, primarily publishing, research evaluation and funding, open science and data sharing. Next, a section will consider research communication, the role of Wikipedia and social media, as well as the online presence of scientists during the crisis. Eventually, the impact of COVID-19 on researchers will be discussed, in particular with respect to gender, inclusion and access to opportunities, disparate impact on career trajectories, mental health. In closing, the chapter will offer some thoughts on the question of what will be the long-lasting impact of COVID-19 on research.
\end{abstract}

Keywords (up to 6): COVID-19, open science, machine learning, research practices, science communication, gender and career trajectories.

\section*{Introduction}

The contribution of research to overcome COVID-19 has been significant. At the same time, the pandemic had and is still having a major impact on research and researchers. meta-research contributions on the matter abound. After more than one year since the start of the pandemic, this review focuses on early meta-research work on understanding how science reacted to COVID-19, as well as on the impact COVID-19 had on research and researchers. The following topics are considered:
\begin{enumerate}
    \item Mapping COVID-19 research and research collaboration during the early stages of the pandemic.
    \item COVID-19 research as data, its use for machine learning applications.
    \item The impact of COVID-19 on research practices including open access and open data, reviewing and publishing, funding.
    \item Science communication beyond peers and towards the general public, including altmetrics and Wikipedia.
    \item The impact of COVID-19 on researchers, in particular with respect to disparate gender impact and career trajectories.
\end{enumerate}
An earlier overview of the impact of COVID-19 on research offers a useful comparison, in particular from the perspective of clinical medicine~\cite{sohrabi_impact_2021}.

The present overview cannot include all possible contributions related to meta-research on COVID-19. Instead, it provides for an overview of the early trends which have started to emerge from the literature. The reader will immediately appreciate how much work lays ahead and how several questions remain largely unanswered, in particular with respect to the long-term consequences of COVID-19 for scientific research. At the same time, this review finds clear patterns in the literature and showcases the meta-research rapid response to COVID-19.

\subsection*{Methods and scope}

The literature considered for this review has been assembled in three steps.

The author selected a first set of impactful research contributions on the topics of the review, using keyword searches on Google Scholar and his expertise. A contribution, in this context, includes full articles, reviews and surveys, (working) research documents and reports, chapters, letters, either peer-reviewed or not (e.g., pre-prints), but not commentaries, viewpoints, position papers, editorials, news items, lectures and presentations or other forms of related outputs which do not contain novel results. Sometimes, such contributions might still be cited to provide context. What constitutes an ``impactful'' contribution depends on the discipline, publication venue, authors, citation counts and the time of publication. For example, citation counts have been taken into account more for older publications, while publication venue and authors have been taken into account more for recent publications, as a proxy for their present and future impact. The selection is particularly difficult for recent pre-prints, which have been included only when they make a clearly novel and complementary contribution with respect to other related and peer-reviewed work. Only contributions released in 2020 or 2021 are considered. These selection criteria have been applied consistently in all subsequent steps as well.

Next, the author systematically explored citation links, both in-citations and out-citations, to expand the initial set of contributions, following the same criteria. Contributions found via citation links were explored in their one-step citation networks accordingly as well. This process continued until no further contributions of interest were found. Google Scholar was again used for this purpose. Thirdly, the author sifted through all contributions mentioning the keyword ``covid'' in their title or abstract, for a selection of quantitative\footnote{Scientometrics, Journal of the Association for Information Science and Technology, Journal of Informetrics, Quantitative Science Studies, Frontiers in Research Metrics \& Analytics.} and qualitative\footnote{Research Evaluation, Minerva, Accountability in Research, Science and Public Policy, Science, Technology, \& Human Values.} science studies journals, as well as general multidisciplinary journals\footnote{Nature, Science, Proceedings of the National Academy of Science, Nature Communications, Scientific Reports, Science Advances, PLoS ONE, eLife, F1000Research, PeerJ, Royal Society Open Science.}. This sifting was done using Dimensions~\cite{hook_dimensions_2018}.~\footnote{The last date when this sift was performed was May 6, 2021. Publications indexed in Dimensions after this date are therefore not part of the review. The reference list reports the month and year for each cited publication.} In the case of mega-journals, the author used Field of Research~\footnote{\url{https://app.dimensions.ai/browse/categories/publication/for}.} categories to only sift through contributions \textit{not} included in the Biological Sciences (division 06), Agriculture and Veterinary Sciences (division 07) or Medical and Health Sciences (division 11).

At all stages, manifestly opinionated pieces which assume or strongly support personal, political or ideological views, even when published as research contributions, have not been included. Some exceptions have been made when substantially novel results are still brought forward. In such cases, their discussion here only focuses on results and not on the personal views of the authors. The resulting selection is necessarily influenced by the author's expertise and critical choices, not to mention possible inadvertent omissions. Nevertheless, it is hoped that a reliable overview of early work on the impact of COVID-19 on research and researchers has been provided, and shall at least serve as a starting point for future work.

\section*{Mapping early COVID-19 research}

Research maps are an important tool of meta-research~\cite{chen_science_2017}. They allow to understand the volume, pace and typology of contributions, their thematic organization, as well as patterns of collaboration within and across teams, institutions and countries. Many contributions have mapped COVID-19 research, in particular during the early weeks and months of the pandemic, also in comparison with previous epidemic outbreaks. This section provides an overview of their approaches and findings.

\subsection*{Tracking research}

One of the meta-research community's most rapid reactions to the COVID-19 pandemic has been to map research efforts related to it. A wealth of studies focused on quantifying the sheer volume of new contributions, and their geographical provenance which follow the diffusion of the pandemic. These contributions are also characterized by a varied use of data sources, which makes them difficult to compare. The earliest publications on COVID-19 up until March 2020 included came primarily from China and other Asian countries~\cite{chahrour_bibliometric_2020} (using PubMed and the WHO database), followed rapidly by the US~\cite{dehghanbanadaki_bibliometric_2020} (using Scopus) and EU countries~\cite{tran_studies_2020} (using the Web of Science, Medline and Scopus), in particular the most severely hit ones (Italy, UK)~\cite{martinez-perez_citation_2020}. Some studies could already spot variations in research focus according to the specific challenges facing different countries~\cite{tran_studies_2020}. Eventually, several authors provided domain-specific early maps of research on COVID-19, on topics as varied as safety~\cite{haghani_scientific_2020} or business and management~\cite{verma_investigating_2020}. 

The first early comparison of COVID-19 research coverage across bibliometric databases, up to mid April 2020, found that Dimensions and Google Scholar seemed to provide the best recall also outside of the biomedical domain~\cite{kousha_covid-19_2020}. This result broadly aligns with more general overviews of bibliometric databases~\cite{martin-martin_google_2021,visser_large-scale_2021}.

Another important trend which emerges in these early mappings relates to the publication typologies of early COVID-19 research. An overview of the typology of early studies in PubMed found significant increases in narrative reviews and expert opinions, followed by case series and reports~\cite{jones_evaluating_2020}. The authors also found that meta-analyses, systematic reviews, and randomized controlled trials remained the least represented publication typology, signaling a lack of evidence-based contributions. These results are echoed by other studies, also reporting an abundance of opinion pieces and a relative lack of novel, evidence-based contributions during the early weeks of the pandemic~\cite{di_girolamo_characteristics_2020,gianola_characteristics_2020,zyoud_mapping_2020}. Arguably, since evidence-based studies take more time to conduct and publish than opinion pieces, it is possible that future work will find that this early trend is at least partially a self-correcting one.

\subsection*{Research topics}

The study of the volume, typology and provenance of publications has been complemented by work focused on uncovering the topics of early COVID-19 research, primarily using text mining and citation network analysis. 

The use of text mining techniques, ranging from keyword analysis to topic modelling, rapidly allowed to establish that health and medical contributions dominate early COVID-19 work. Three clear broad themes emerge in this literature: virology, emergency care and public health, as well as contributions on the global and local responses to the pandemic~\cite{tran_studies_2020}. While the biomedical disciplines were understandably leading on early COVID-19 research contributions, the expansive trend of COVID-19 research could already be noticed before mid-2020 with other disciplines such as those in the social sciences and humanities joining the effort~\cite{aristovnik_bibliometric_2020}. Further examples of the use of machine learning and natural language processing to map early COVID-19-research abound. \cite{colavizza_scientometric_2021} uses topic modelling on the July 1, 2020 version of CORD-19~\cite{wang_cord-19_2020} (see below) to find a clear focus on few research topics including coronaviruses (primarily SARS-CoV, MERS-CoV and SARS-CoV-2), public health and viral epidemics, the molecular biology of viruses, influenza and other families of viruses, immunology and antivirals, clinical medicine. Useful comparisons to this study, yielding similar results, can be found for the very same version of CORD-19~\cite{abd-alrazaq_comprehensive_2021} and for other data sources as well~\cite{ebadi_understanding_2021}.

Citation network analyses of the early COVID-19 literature offer a similar picture, showcasing a clear rise of research on public health, including COVID-19 treatment and epidemiology, as well as its psychological impact~\cite{martinez-perez_citation_2020}, and including analyses of temporal trends~\cite{furstenau_bibliometric_2021}.
Comparisons with previous epidemics, including SARS and MERS, found that all three outbreaks generated similarly distinct cohorts of studies: public health responses and epidemic control; virology; treatment, vaccines and clinical care~\cite{haghani_covid-19_2020}. Furthermore, COVID-19 research appeared since early on to not only be much larger in volume, but also broader and more driven by international collaboration and geographical patterns related to when and how the pandemic unfolded~\cite{pal_visualizing_2021}, although such results are contested~\cite{digital_science_how_2020}. Importantly, research on COVID-19 shows less cohesion and more new directions emerging when compared with previous, more localized outbreaks. Epidemiology research appears to be the most stable topic before and during COVID-19~\cite{zhang_topic_2021}.

\subsection*{Patterns of collaboration}

A substantial number of publications has focused on the geography of COVID-19 research, including from the point of view of collaborations. Other studies have considered different scales of collaboration, down to the laboratory or research group.

A useful starting point is a comparison of early COVID-19 research with five other disease outbreaks since 2000~\cite{zhang_how_2020}. The authors show that ``academia always responded quickly to public health emergencies with a sharp increase in the number of publications immediately following the declaration of an outbreak by the WHO.'' While most countries are primarily concerned with epidemics impacting their own region, European and North American countries tend to have a global focus and conduct collaborative research with the countries concerned with an outbreak, such was for example the case of Ebola in Africa. A substantial international degree of collaboration, the dominance of big research players (US, China, EU, UK, India), and consistent open access availability have been amply confirmed for COVID-19 research since~\cite{lee_scientific_2021,belli_coronavirus_2020}. Given the global scope of COVID-19, these patterns do not surprise~\cite{li_analyzing_2021}. Indeed, a study of mainstream medical journals found that the geographical origins of COVID-19 authors break down as follows: European (47.7\%), North-American (37.3\%), China (8.8\%) and the rest. While these results align with general research publication patterns, including in times of epidemic outbreak, they also showcase the persisting imbalances in research capacity across countries and regions~\cite{benjamens_are_2021}.

Patterns of collaboration at the team scale have also been considered. A narrowing of team membership has been spotted in early COVID-19 research, as well as an even greater visibility of elite institutions~\cite{homolak_preliminary_2020}. Both the size and the location of teams in elite institutions might be related to access to resources. This is further evidenced by fewer than expected acknowledgements of funders in publications from the early months of the pandemic, when compared to pre-pandemic research. The teams more readily able to respond to the outbreak have been those with funding already in place, or no need for funding at all, and they acted in smaller, agile groups at first~\cite{fry_consolidation_2020}. The same authors have found all these trends to persist or even increase in a closely-related followup after a few months~\cite{cai_international_2021}.

\subsection*{Outlook}

The impact of these abundant efforts into mapping early COVID-19-related research remains unclear. After an initial enthusiasm, maps of research seem to have been less popular after the initial months of the pandemic, as it became clear that the magnitude of the event was gaining attention from most fields of science. At the same time, early maps of research might have helped to track the immediate research reactions to the pandemic, while specialized tools and search engines were not yet available for this purpose. Global retrospective maps of COVID-19-related research have still to be produced and fully analysed, while more topically localised maps could still be useful to serve the specific needs of a given community.

The early research response to the COVID-19 pandemic also shows clear signs of the enduring strengths and weaknesses in global research capacity. The US, EU, UK, China and India led the early and subsequent research efforts, acting as hubs for international collaboration, while other countries seemed more focused on local issues. At a small, team scale too, being able to leverage existing resources has been key in allowing elite groups and institutions, in particular, to move quicker and earlier. This evidence underlines, yet again, the importance of an existing, long-term research capacity in rapidly and effectively reacting to a crisis.

Substantive questions in terms of global research trends on COVID-19 remain open and are only starting to be approached by the community. For example, \cite{ioannidis_rapid_2020} considers the rapidity and scale at which authors from any field of research shifted their focus on COVID-19 topics during the first full year of pandemic -- using Scopus, up to March 1, 2021. They find that all fields of research contributed to COVID-19 research, with half a million researchers having published something on COVID-19 and 98 over 174 sub-fields having at least one influential author publishing on COVID-19. Such massive shift of focus is unprecedented, the authors in fact refer to a ``covidization'' of research including cases of ``hyper-productivity'', whose implications will have to be monitored and are largely to be understood.

\section*{Data and machine learning}

Health crises are also information crises~\cite{xie_global_2020}. COVID-19 has accelerated the existing trend in making scientific literature available as data, to be mined and analysed using techniques from data science and machine learning. Furthermore, it has fostered the development of tools to navigate and make rapid sense of it, such as search engines or question answering systems.

\subsection*{COVID-19 research as data}

Several initiatives have focused on providing rapid access to COVID-19 literature as data, sometimes also via graphical user interfaces. Collections of scholarly articles on COVID-19 are one of the main related open source contributions which have been made~\cite{shuja_covid-19_2021}. Notable initiatives include LitCOVID, CORD-19, and Dimensions~\cite{hook_real-time_2021}. 
LitCOVID~\cite{chen_litcovid_2021} is innovative in that it extensively applies machine learning to select COVID-19 literature from PubMed, extracting topics, geolocations and other keywords. The dataset is updated daily and free to download.
The COVID-19 Open Research Dataset (CORD-19)~\cite{wang_cord-19_2020}, developed by The Allen Institute for Artificial Intelligence and steadily improved over time~\cite{rohatgi_covidseer_2020,kanakia_mitigating_2020,colavizza_scientometric_2021}, has become the de facto standard for text-mining tools on COVID-19 literature. This dataset gathers publications from a variety of sources, including PubMed and pre-print servers, and equips them with metadata and the full-text, when available. This dataset is also updated daily and free to download.
Another related and high-potential innovation are knowledge graphs, which offer a machine-readable and expressive way to represent scholarly literature and its contents and constitute an emerging area of active research which promises to be rapidly integrated into science communication tools in the near future~\cite{jaradeh_open_2019}. Several contributions extracted knowledge graphs from scientific literature on COVID-19~\cite{domingo-fernandez_covid-19_2020,wise_covid-19_2020,cernile_network_2021,reese_kg-covid-19_2021,wang_covid-19_2021}. While these experiments might need more time to find their way into systematic usage, they are promising and have been substantially accelerated by COVID-19. See \cite{chatterjee_knowledge_2021} for a recent overview.

\subsection*{Machine learning applications}

The importance of machine learning and data science in fighting COVID-19 has been recognized and surveyed early on~\cite{bullock_mapping_2020,tayarani_n_applications_2021,shorten_deep_2021}. This also applies to science communication and the need to make sense of a very rapidly growing body of literature. In particular, several search engines have been developed since the very early stages of the pandemic~\cite{kricka_artificial_2020}. Overviews of these tools and their use to follow topics of COVID-19 research are also available~\cite{porter_tracking_2020,mercatelli_web_2021}. Given the recent rapid growth of machine learning and data science, and their current importance for many research and industry tools, it perhaps does not come as a surprise that COVID-19 ushered an abundance of attempts to use them in order to make sense and contribute to the research effort.

Many machine learning contributions have relied on CORD-19, and derived datasets have been created. Examples include entity recognition and concept extraction~\cite{wang_comprehensive_2020,kroll_semantically_2020,cernile_network_2021,reese_kg-covid-19_2021,wang_covid-19_2021}, question answering~\cite{tang_rapidly_2020}, textual evidence mining~\cite{wang_automatic_2020}, text similarity~\cite{guo_cord19sts_2020,wise_covid-19_2020}, search engines~\cite{zhang_rapidly_2020,macavaney_sledge-z_2020,zhang_covidex_2020,esteva_covid-19_2021}, graphical interfaces to explore scientific evidence on COVID-19~\cite{verspoor_covid-see_2020,hope_scisight_2020}, keyword extraction and clustering~\cite{doanvo_machine_2020}, summarization~\cite{kieuvongngam_automatic_2020}. Text mining approaches to COVID-19 literature have been critically assessed in a recent review~\cite{wang_text_2021}.

Following an established approach in machine learning research, special conference tracks and shared tasks have been set-up to mine and enrich COVID-19 literature. A special TREC COVID track was set-up to test search engines for COVID-19 literature using CORD-19, and to develop a test collection of searches for assessing future systems in this respect~\cite{roberts_trec-covid_2020}. The track has been extremely popular and achieved its stated goals~\cite{roberts_searching_2021}, as shown by the continued use of their data for evaluation (e.g., \cite{esteva_covid-19_2021}). It is important to note that significant limitations might exist in such academic prototypes when compared to industry search engines~\cite{soni_evaluation_2021}.
Other shared tasks include the Kaggle CORD-19 research challenge~\footnote{\url{https://www.kaggle.com/allen-institute-for-ai/CORD-19-research-challenge}.} and the epidemic question answering challenge~\footnote{\url{https://bionlp.nlm.nih.gov/epic_qa}.}. Further initiatives included a workshop on Natural Language Processing for COVID-19 at the 2020 Annual Meeting of the Association for Computational Linguistics (ACL)~\cite{verspoor_proceedings_2020} and its followup at the Empirical Methods in NLP conference (EMNLP)~\cite{verspoor_proceedings_2020-1}.

\subsection*{Outlook}

Data science and machine learning contributions to COVID-19 require, first of all, the availability of scientific literature as data to be mined. In this respect, COVID-19 has shown that well-maintained, well-resourced and regularly updated initiatives fare better, making a longer-lasting impact. Furthermore, widespread institutional backing and an openness to talk to and be informed by domain users also play a crucial role in fostering the adoption and improvement of a dataset. A clear exemplar in this respect is CORD-19.

During the first year of the pandemic, a variety of machine learning and other data-driven methods have been used. It is likely that most of them will not make a direct lasting impact, yet they signal and accelerate trends that were already in place before COVID-19. Search engines such as the Semantic Scholar, also from the Allen Institute, since quite some time integrate machine learning services to support the search, retrieval and understanding of scientific literature~\cite{ammar_construction_2018,cachola_tldr_2020,beltagy_scibert_2019,cohan_specter_2020}. Advances in making the scientific literature machine readable were ongoing too~\cite{jaradeh_open_2019}. A systematic and extensive use of machine learning and other data-driven methods, such as knowledge graphs, clearly constitutes a next frontier in science communication. COVID-19 has made their necessity clear, and also likely accelerated the pace of their experimentation, if not adoption. 
%It has also made manifest how insufficient the current science communication ecosystem is in dealing with the scale of ever new scientific contributions. Scientific knowledge has to be made machine-readable if a new generation of applications is to be developed in order to deal with such staggering abundance.

\section*{Research practices}

COVID-19 has not only impacted what scientists focus on, but also how they operate. This section reviews meta-research on open access and open data, reviewing, publishing and funding early COVID-19 research.

\subsection*{Open access and open data}

The pandemic has prompted researchers and many publishers to rapidly adopt radical open access initiatives, which made most related research openly available. In this way, most (more than 90\%) early COVID-19 literature was already made available in open access during the early stages of the pandemic~\cite{belli_coronavirus_2020}. At the same time, the sheer volume of publications likely overloaded the traditional publication system and led to issues with data and code availability for those same early publications~\cite{homolak_preliminary_2020}. A relatively low degree of data sharing in early COVID-19 research has indeed been confirmed looking at PubMed Central publications. Research found that only about 28.5\% articles provided at least one link to a dataset~\cite{zuo_how_2021}, despite efforts in this direction\footnote{\url{https://wellcome.org/coronavirus-covid-19/open-data}.}. It appears that, while many publishers opened up their publications, the adoption of open science principles and practices would have avoided many issues in early COVID-19 research, including ``research waste'' (repeated studies), dubious quality and even retractions~\cite{besancon_open_2020}.

\subsection*{Reviewing and publishing}

The use of pre-prints to share COVID-19 research has been widespread, similarly to their unprecedented and sometimes non-transparent relay by news media~\cite{fleerackers_communicating_2021}. COVID-19 pre-prints are on average shorter, reviewed faster, and receive more engagement, including from news media and the public, than non-COVID-19 pre-prints~\cite{fraser_evolving_2021}. Importantly, a study found that only a small fraction of early pre-prints were eventually published as journal articles (5.7\%), despite accelerated review times. Those eventually published also showcased higher citation rates on average~\cite{anazco_publication_2021}. These results should be taken with caution given that the quality of the linkage between pre-prints and journal articles remains poorly understood~\cite{lachapelle_covid-19_2020,cabanac_day--day_2021}.

The rapid adoption of pre-prints clearly signals a need for speed which was not met by the scientific publishing system. The abundance of COVID-19 literature, as well as the pressure to publish rapidly, might also have led to more lenient peer-review practices. A qualitative assessment of COVID-19 peer-review for medical journals found that ``no clear differences between the review processes of articles not related to Covid-19 published during or before the pandemic. However, it does find notable diversity between Covid-19 and non-Covid-19-related articles, including fewer requests for additional experiments, more cooperative comments, and different suggestions to address too strong claims''~\cite{horbach_no_2021}. Several concrete suggestions on how to improve peer review practices have been advanced~\cite{teixeira_da_silva_optimizing_2021}. A variety of initiatives have since started to speed reviews up while attempting not to compromise on quality. An example is the journal eLife, which adopted pre-prints by default, no more experiments asked nor deadlines in revisions, mobilized junior academics and attempted to smooth competition for priority~\cite{eisen_publishing_2020}. Another example is the rapid review initiative~\cite{hurst_covid19_2021}. Innovative forms of reviewing have also been used, for example via crowdsourcing on pre-prints. Examples include the Outbreak Science PREreview\footnote{\url{https://outbreaksci.prereview.org}.} and the MIT Press Rapid Reviews COVID-19\footnote{\url{https://rapidreviewscovid19.mitpress.mit.edu}.}

Despite clear successes in accelerating the speed of peer reviews, the deluge of COVID-19 articles was soon coupled with warnings of how rapid acceptance speeds might also come at the detriment of integrity~\cite{dinis-oliveira_covid-19_2020,glasziou_waste_2020,palayew_pandemic_2020,horbach_pandemic_2020,bramstedt_carnage_2020}. The relatively high rate of retractions in early COVID-19 research was also found to be a possible consequence of the speed and sometimes carelessness in publishing~\cite{soltani_retracted_2020,yeo-teh_alarming_2021}. The lack of research data integrity has been suggested as another aggravating factor in this respect~\cite{shamoo_validate_2020} and, eventually, predatory publishing practices were also spotted~\cite{teixeira_da_silva_alert_2020}. Nevertheless, other authors have cautioned to rush to conclusions until more systematic studies become available~\cite{abritis_alarming_2021}.

\subsection*{Research funding}

Funding agencies rapidly reacted to COVID-19 with measures to mitigate its impact, including extensions of deadlines and dedicated support~\cite{stoye_how_2020}. At the same time, many countries have by now released dedicated funding instruments for COVID-19 which are still to be mapped out and assessed. It is worth noting that several communities, in particular during the early months of the pandemic, stressed the importance of long-term, stable funding over that of reactive, short-term funding~\cite{prudencio_research_2020}. The importance of shifting focus to the long-term consequences of COVID-19, in particular from an interdisciplinary perspective, have also been clearly stressed~\cite{yelin_long-term_2020}, notwithstanding the need to meet immediate priorities~\cite{zeggini_biomedical_2020}. Indeed, public and long-term funding is key in delivery preparedness such as in the case of the development of the Oxford-AstraZeneca COVID-19 vaccine~\cite{cross_who_2021}. The authors of this study found that nearly all funding related to the development of this vaccine came from public sources, since at least 2002. Furthermore, the authors ``encountered a severe lack of transparency in research funding reporting mechanisms.'' It is doubtful whether COVID-19 rapidly deployed instruments will fare much better in this respect.

\subsection*{Outlook}

The early research response to COVID-19 has come as a flurry of results, often released and re-used as open pre-prints. The importance to find solutions, and share them openly, has been met in this way by many. The shortcomings of the traditional publishing system, with its slow procedures, have become clearly visible during the pandemic. Nevertheless, the rapid release of research likely had negative consequences as well, including poor open science practices, possibly lower quality standards and a higher rate of retractions, a proliferation of superficial or duplicated results. In terms of research funding, while understanding what has been done and its impact remains an open question, there seems to be agreement on the need to shift focus from short-term projects to understanding the long-term impact of COVID-19. It also appears clear that scientific excellence and preparedness cannot be summoned overnight but require structural planning and resourcing.

\section*{Communicating research}

COVID-19 has put science and scientists on the spotlight. The search for a vaccine, the evidence for public policy measures, and more recently discussions on their impact are among the topics which attract wide public interest. Furthermore, conveying science-backed information to the public has been key, in particular to mitigate misinformation, as is the case of Wikipedia. Lastly, social media such as Twitter play an increasingly important role in sharing and discussing results among peers and more broadly.

\subsection*{Altmetrics and Wikipedia}

While the diffusion of trusty and untrustworthy information on social media has been a clear theme of COVID-19~\cite{cinelli_covid-19_2020}, the same social media can also provide a venue to rapidly share scientific results and identify noteworthy ones via altmetrics~\cite{sugimoto_scholarly_2017}. Indeed, there have been proposals to use altmetrics to identify impactful COVID-19 research since early in the pandemic~\cite{boetto_using_2021}. Similarly, another application of altmetrics might be to detect early signs of criticism leading to a retraction~\cite{haunschild_can_2021}. In fact, initial analyses have shown that COVID-19 research gathers higher than average mentions on Twitter, also thanks to the dissemination efforts of scientists~\cite{fang_tracking_2020}. Other work has also found evidence for a correlation between altmetrics, such as Twitter mentions, and later citation counts~\cite{kousha_covid-19_2020}. While it seems premature to draw general conclusions, the heavy use of social media in science communication is definitely a significant aspect of the COVID-19 pandemic, which warrants more attention.

Wikipedia is another source of reliable and trustworthy information on the Web~\cite{singer_why_2017,lemmerich_why_2019}. Wikipedia showed a remarkable surge in editorial activity and onboarding of new editors during the early stages of the pandemic, with a 20\% increase in contributions for English Wikipedia, when compared to a pre-pandemic baseline~\cite{ruprechter_volunteer_2021}. These contributions have been particularly significant in creating new or updating existing COVID-19-related articles~\cite{keegan_quantitative_2020}. In this respect, new COVID-19 contents added to Wikipedia during the early months of the pandemic have usually relied on recent, highly-visible, open access and peer-reviewed scientific sources~\cite{colavizza_covid-19_2020}, or trustworthy news media sources~\cite{benjakob_meta-research_2021}. Notably, similar high-bar inclusion criteria have been found to hold for policy documents, particularly those released by inter-governmental organizations (IGOs), such as the World Health Organization (WHO)~\cite{yin_coevolution_2021}.

\subsection*{Scientists and the public}

An early and visible contribution discussed an array of insights and open challenges related to COVID-19 from the perspective of social and behavioural sciences~\cite{bavel_using_2020}. The main challenges associated with science communication as identified by the authors were: conspiracy theories, fake news and misinformation, persuasion. These themes are now new~\cite{west_misinformation_2021}, yet they came to the fore with COVID-19. Indeed, some studies have considered the public role of scientists during COVID-19. For example, a US-focused survey found no evidence of the alleged negative impact that President Trump's negative remarks might have had on scientists' perceived trustworthiness~\cite{evans_who_2020}. Instead, the authors found a broad public agreement on scientists’ claims about COVID-19, but more resistance and disagreement when descriptions turn into prescriptions and value-laden policy recommendations. Social media, and in particular Twitter, have also offered an important venue for scientists to reach the public. Crucially, evidence seems to show that while false information on COVID-19 in generally tweeted more, science-based evidence and fact-checking tweets receive capture more engagement~\cite{pulido_covid-19_2020}. The seemingly positive role of scientists in conveying trust, or the spokesperson effect, has found evidence in the case of COVID-19 information. A large-scale survey found that ``across countries and demographic strata, immunology expert Dr. Anthony Fauci achieved the highest level of respondents’ willingness to reshare a call to social distancing, followed by a government spokesperson. Celebrity spokespersons were least effective''~\cite{abu-akel_effect_2021}. Similar results have also found independent confirmation~\cite{farjam_dangerous_2021}.

\subsection*{Outlook}

The research on COVID-19 has rapidly become highly visible on online and social media. Similarly, scientists appear to have been more active in sharing their work and thoughts. On average, there are early positive signs of the impact of such activities. Social media showed promise to help identify impactful COVID-19 research, while Wikipedia appears to have stood up to the challenge of updating its contents with reliable evidence, as it became available. Lastly, scientists have been found able to play a significant role as spokespersons for trustworthy information. Taken together, these signs underline the increasing importance and potential of social and online media in science communication, in particular during times of crises when science is asked to provide answers and solutions. At the same time, they underline how little we know on the mechanics of science communication beyond peers. Finally, it must be noted that other important aspects of science communication, namely to medical professionals and to policy makers, have not been considered in this section.

\section*{The impact of COVID-19 on researchers}

The impact that COVID-19 had on researchers and research is hard to understate. Questions abound, including understanding the possibly disparate impact of COVID-19 on different groups and individuals, on career trajectories, on the mental health of researchers and students, on teaching and virtual events, and much more. Nevertheless, caution should be applied as short-term trends might not necessarily be indicative of substantial long-term consequences.

\subsection*{Gender and caring responsibilities}

Several early and descriptive studies rapidly and alarmingly called for a possible under-representation of women in COVID-19 research. A lower contribution of women as first authors in medical journals during the very first months of the pandemic has indeed been individuated, which apparently was not present for women senior authors~\cite{andersen_covid-19_2020}. Another study took this approach one step further by using differences-in-differences to estimate the impact of lock-downs on general research productivity~\cite{cui_gender_2020}. Controlling for discipline, career stage and institutions, as well as gender, the authors found that the only significant effects of COVID-19 lock-downs on research productivity, during the first 10 weeks of the pandemic, concentrated on assistant professors, in particular at top-rated institutions. This suggest that higher-pressures and a possible gender imbalance in caring responsibilities (e.g., child-caring), in aggregate, could have led to a disparate impact on research productivity for some women. These points are supported by other studies, focused on the early impact of the pandemic on (academically) younger women~\cite{squazzoni_no_2020}, and with a particular severe drop of women's contribution to COVID-19 biomedical research~\cite{muric_gender_2021}. Caution should still be applied, as country-level results in this respect appear to differ markedly~\cite{abramo_gendered_2021}, while some domain-specific studies even found an overall higher women productivity in 2020 than 2019, when considering non-COVID-19-related research, as shown for cardiovascular journals~\cite{defilippis_gender_2021}. Furthermore, there are since indications that the reduced productivity of young women on COVID-19 research is a resolving trend~\cite{lerchenmuller_longitudinal_2021}.

The connection among research productivity, parenting and other caring responsibilities has been further explored with survey-based methods. Despite their misleading title, \cite{deryugina_covid-19_2021} show that being a parent and, in particular, a woman parent, often entails substantial reductions in research productivity, as compared with non-parents of either sex.
%\footnote{Despite these clear results, the authors of this work prefer to conclude their study by stating: ``Our time-use survey suggests that the short-term adverse productivity effects of the pandemic fall disproportionately on female academics with children.'' Furthermore, they title it as follows: ``COVID-19 disruptions disproportionately affect female academics''. It remains inexplicable why, from finding a disproportionate impact on parents in general (of both sexes), the authors decide to generalise their results like they did.}
Similar results have been amply supported~\cite{yildirim_differential_2021,krukowski_academic_2021,myers_unequal_2020}, and have been confirmed outside of the academic context as well~\cite{giurge_multicountry_2021,hipp_parenthood_2021}. The disparate impact of parenthood in academia is known more generally, and it appears to explain a significant share of the gender productivity gap~\cite{morgan_unequal_2021}, including by higher drop-out rates and different career lengths for young academics with caring responsibilities~\cite{huang_historical_2020}. This trend, in particular when considering mothers, also appears to be shrinking over time.

Gendered funding policies have already been put in place~\cite{witteman_covid-19_2021}. Such measures do not seem to be justified given the available evidence, which would if anything support interventions targeted at supporting young academics (of any gender) with caring responsibilities. Nevertheless, more systematic studies on the longer-term impact of COVID-19 on researchers from different demographic groups, backgrounds and career trajectories are necessary, as is the constant monitoring of the situation, in order to potentially inform policy interventions.

\subsection*{Other challenges}

While this section cannot do justice to all the multifaceted ways in which COVID-19 is impacting researchers, a few emerging trends can already be mentioned. While rapidly shifting to working from home has been a complex and at times traumatic experience, recent work suggests that many academics might prefer to keep more flexibility in this respect in the future~\cite{aczel_researchers_2021}. Indeed, some recent changes in work arrangements, such as remote meetings, might stay in the post-COVID academic life. Another example in this respect are conferences and other academic events. COVID-19 has, if anything, called for a reflection on the costs (e.g., for the environment) of the academic lifestyle, as well as made it clear that virtual options are viable in this respect. Conferences thus face both challenges and opportunities in a post-COVID academic world~\cite{mubin_new_2021}. Less optimistically, it is likely that COVID-19 has exerted a heavy toll on researchers mental health, physical and psychological well-being. Early evidence finds higher burnout and stress levels, in particular for the medical disciplines involved with COVID-19 patients~\cite{odriozola-gonzalez_psychological_2020,kannampallil_exposure_2020,fornili_psychological_2021}. The impact of COVID-19 on students will also be a significant area of future work. Early research suggested a possible impact on their psychological well-being~\cite{browning_psychological_2021,elmer_students_2020} and on learning~\cite{aristovnik_impacts_2020,gonzalez_influence_2020}, not necessarily all negative. This finds echo in primary and secondary schools as well~\cite{engzell_learning_2021}.

\subsection*{Outlook}

The impact of COVID-19 on researchers and students remains largely to be understood. Early research seems to suggest that, at least during the early stages of the pandemic, the shift to virtual and the lock-downs have negatively impacted the productivity of younger academics with caring responsibilities, primarily parents. Such finding seems largely, albeit not completely, to explain a possible disparate impact on women as well. Several other important topics have at least started to emerge in early meta-research on COVID-19, including discussions on the future of academic work and events in a post-COVID world, the psychological and physical consequences of the pandemic, the impact of restrictions due to COVID-19 on teaching and education more broadly. All these topics, and more, constitute important areas for future research.

\section*{Conclusions}

This review has provided an overview of early COVID-19-related meta-research, after little more than one year from the beginning of the pandemic (including contributions released until May 6, 2021). Some trends emerge, which can be summarized as follows:

\begin{itemize}
    \item COVID-19 determined a rapid and massive shift of attention by researchers in most disciplines. This is understandable as COVID-19 has impacted so many aspects of life, yet it remains to be seen how stable the shift will actually be. In particular, as research on COVID-19 will move from a focus on the short to the long-term impact of COVID-19, a partial attention re-balancing might occur.
    \item The availability of scholarly literature as data generated a surge in experiments with machine learning techniques to facilitate research communication and understanding. While most of these results will likely not find immediate use, they constitute an acceleration of a trend pre-dating COVID-19. Success stories such as CORD-19, further exemplify how a combination of resource availability, institutional support, and continued maintenance are important ingredients in not only creating but also establishing a successful data resource or machine learning application.
    \item Science communication has been significantly impacted by COVID-19. The need for rapidly sharing results fostered the adoption of open access, the use of pre-prints and of accelerated and sometimes innovative reviewing procedures. Nevertheless, traditional practices such as peer-review and grant-based funding have been visibly challenged by COVID-19, in particular because they often could not cope with the rapid action which was required. At the same time, and partially as a consequence, a deluge of low quality contributions was produced. While peer-review and funding systems seem likely to withstand COVID-19, the growing use of openly accessible pre-prints might become increasingly more common after it.
    \item Science communication went increasingly social with COVID-19. The pandemic gave some scientists a prominent public role on both traditional and social media. COVID-19 made it clear that science and scientists do not exist in a vacuum, but instead they increasingly have public visibility. Social media-based altmetrics, such as Twitter mentions, have also been found to correlate with later citation impact for COVID-19 literature: a novel trend which will require further monitoring. Lastly, the pandemic also put to the test the resilience and responsiveness of open commons such as Wikipedia, which responded admirably.
    \item COVID-19 has had a significant and still largely to be understood impact on researchers. Worrying signals of gender-based disparate impact appear to be primarily related to caring responsibilities -- such as parenting -- and the high pressures on young academics due to hyper competition and lack of job safety, in particular at elite institutions. The future of academic work also remains open for change, as some virtual options might be here to stay.
\end{itemize}

The most important questions related to the impact of COVID-19 on research and researchers remain open. This review highlighted how early meta-research on COVID-19 reacted to short-term questions and challenges, and has only recently started to more systematically explore the longer-term consequences of COVID-19. Indeed, over the coming years it will be crucial to shift attention from spotting short-term trends to understanding the long-term consequences of COVID-19 for science, broadly construed. While it is impossible to overstate the grim impact that COVID-19 had and is having on society, it can also be an opportunity to cast new light on many outstanding questions in the social sciences and beyond~\cite{conley_opinion_2021}. Perhaps this crisis shall serve as a catalyst to reflect and improve on how we do science too.

\subsection*{Acknowledgements}

The author would like to thank Ludo Waltman (CWTS, Leiden University) for helpful comments and suggestions on an earlier draft.

\bibliography{bibliography}
\end{document}